\documentclass[conference]{IEEEtran}
\IEEEoverridecommandlockouts

\usepackage{cite}
\usepackage{amsmath,amssymb,amsfonts}
\usepackage{algorithmic}
\usepackage{graphicx}
\usepackage{textcomp}
\usepackage{xcolor}
\usepackage{xspace}
\usepackage{newtxtext}
\usepackage{multirow}

\usepackage{tikz}
\usetikzlibrary{positioning, fit, arrows.meta, calc, shapes.geometric}

\usepackage{xspace}
\usepackage{booktabs}
\usepackage{tabularx}
\usepackage{array}
\usepackage{pifont}
\usepackage{hyperref}

\newcommand{\eg}{e.g.\xspace}

\newcommand{\toolname}{\textsc{CONQuER}\xspace}
\newcommand{\toolopt}{\texttt{conquer-opt}\xspace}

\newcommand{\resultSpeedup}{12.19$\times$\xspace}

\newcommand{\resultAccuracyDrop}{1.44\%\xspace}

\newcommand{\targetMobileCPU}{Snapdragon 8 Elite\xspace}
\newcommand{\targetLaptopCPU}{Intel I5 1145g7\xspace}
\newcommand{\targetServerGPU}{NVIDIA A100 80GB\xspace}

\newcommand{\MLIR}{\texttt{MLIR}\xspace}
\newcommand{\TOSA}{\texttt{TOSA}\xspace}
\newcommand{\Linalg}{\texttt{Linalg}\xspace}
\newcommand{\IREE}{\texttt{IREE}\xspace}
\newcommand{\ONNX}{\texttt{ONNX}\xspace}

\newcommand{\cmark}{\ding{51}}
\newcommand{\xmark}{\ding{55}}

\newcolumntype{L}{>{\raggedright\arraybackslash}X}

\begin{document}

\title{\toolname{}: Hardware-Aware Mixed-Precision Quantisation with Online-Calibrated Surrogates%
\thanks{This work was supported by the Huawei Edinburgh Joint Lab project \emph{RobustCheck: Testing Robustness of Compiler Optimisations and Deep Learning Frameworks}.}
}

\author{\IEEEauthorblockN{1\textsuperscript{st} Aidan Dakhama}
\IEEEauthorblockA{
\textit{School of Informatics} \\
\textit{University of Edinburgh} \\
Edinburgh, United Kingdom \\
adakhama@ed.ac.uk
}
\and
\IEEEauthorblockN{2\textsuperscript{nd} Ajitha Rajan}
\IEEEauthorblockA{
\textit{School of Informatics} \\
\textit{University of Edinburgh} \\
Edinburgh, United Kingdom \\
arajan@ed.ac.uk
}
}


\maketitle

\begin{abstract}
Deploying deep neural networks on resource-constrained hardware relies heavily on mixed-precision quantisation (MPQ). However, current deployment toolchains severely fragment this process. Quantisation typically occurs as a hardware-agnostic preprocessing step in front-end frameworks, disconnected from the downstream compilers that generate the physical machine code. This separation leads to suboptimal configurations where assigned bit-widths map poorly to the target machine's heterogeneous hardware execution blocks such as tensor cores and variable-width vector units, incurring severe runtime execution penalties. Furthermore, evaluating these configurations via exhaustive hardware-in-the-loop (HIL) testing is computationally intractable due to the exponentially large search space.

We present \toolname{}, a unified, compiler-integrated infrastructure for hardware-aware MPQ. \toolname{} shifts quantisation directly into the compiler pipeline at the \MLIR{} \TOSA{} level, enabling intelligent configuration handling based on compiler support. To evaluate this combinatorial search space of different combinations of model layers within practical compilation budgets, \toolname{} couples an NSGA-II evolutionary algorithm with a dual-surrogate pre-screening engine. This engine evaluates theoretical cache memory bounds and feature space isotropy to immediately discard non-viable configurations. \toolname{} then executes only the strongest candidate policies on physical hardware via \IREE{}, feeding the exact execution metrics into a logarithmic online calibrator. This calibrator continuously aligns the surrogate models with the ground-truth hardware behaviour during an NSGA-II evolutionary search.

Evaluation across mobile CPUs, laptop CPUs, and discrete server GPUs demonstrates that optimal quantisation policies are strictly hardware-dependent. By tightly coupling quantisation with compiler lowering and physical execution, \toolname{} discovers Pareto-optimal configurations that achieve up to \resultSpeedup{} faster inference while keeping top-1 accuracy within \resultAccuracyDrop{} of the unquantised \MLIR{} baseline.
\end{abstract}


\section{Introduction}
\label{sec:introduction}

Deep neural networks deliver strong predictive performance, but deploying them efficiently on diverse, resource-constrained hardware remains difficult. Quantisation reduces the precision of weights and activations to lower memory usage and compute cost, and is now a standard part of practical deployment pipelines~\cite{gholami2022survey, jacob2018quantization}. However, applying a uniform bit-width across all layers ignores the fact that different parts of a network exhibit different sensitivity to quantisation noise, often resulting in unnecessary accuracy degradation. 
Mixed-precision quantisation (MPQ) addresses this limitation by assigning different precisions to different parts of the network. 
Additionally, modern AI hardware does not process all maths on a single, uniform arithmetic logic unit (ALU). Instead, chips are split into highly specialised blocks such as tensor cores and matrix engines for hyper-fast low-precision maths, vector units and standard ALUs with FP32 or FP16 precision sitting alongside for more sensitive operations like layer normalisation and activations, while neural processing units support low precision maths operations in edge devices. 
The difficulty with MPQ, however, is no longer whether to quantise, but how to identify an effective policy under real hardware architecture constraints. Evaluating mixed-precision policies is bottlenecked by the deployment toolchain: exhaustively testing configurations for a standard 50-layer network requires years of compilation time. To navigate such massive combinatorial search spaces, evolutionary algorithms like NSGA-II have proven highly effective~\cite{nebro2022nsga}.
Furthermore, the utility of a configuration depends deeply on the physical behaviour of the target hardware.

Recent work has shown that deployment-aware quantisation choices matter. HAQ uses reinforcement learning driven by hardware feedback to search mixed-precision policies for specialised accelerators \cite{wangHAQHardwareAwareAutomated2019}, and HAWQ-V3 formulates mixed-precision allocation as an optimisation problem bounded by hardware-related constraints \cite{yaoHAWQV3DyadicNeural2021}. At the deployment level, SeQTO demonstrates that selective quantisation combined with on-device profiling on a per-model basis can recover substantial accuracy for a given \ONNX{} model \cite{louloudakisSelectiveQuantizationTuner2025}. These works reinforce an important point: realistic deployment metrics must be part of the optimisation loop. However, they also highlight a major practical limitation. Existing approaches either rely entirely on proxy/offline metrics that may correlate only weakly with actual hardware behaviour, or they depend on expensive hardware-in-the-loop (HIL) evaluation for every candidate configuration. In both cases, quantisation remains decoupled from the compilation pipeline, which can lead optimisation to consider mixed-precision configurations unsupported by the target backend.
This gap matters in practice because modern ML deployment increasingly relies on compiler infrastructures such as \MLIR{} to lower high-level models to heterogeneous hardware targets \cite{lattnerMLIRCompilerInfrastructure2020} offering more portability. Frameworks such as \IREE{} already provide a robust lowering and execution pipeline through dialects such as \TOSA{} and \Linalg{} for CPU, GPU, and embedded targets \cite{liuTinyIREEMLExecution2022}. Yet current workflows largely expect quantised models to be supplied by front-end frameworks (e.g., PyTorch, TensorFlow). This early binding creates a rigid dependency, forcing developers to manage separate, hardware-specific quantisation scripts for every framework they use. Shifting MPQ into the compiler's middle-end – specifically at the framework-agnostic \TOSA{} level – offers a highly compatible alternative. However, achieving this without being constrained by the rigid semantics of the \MLIR{} \texttt{quant} dialect remains an open tooling problem.

We present \toolname{}~\footnote{The source code of \toolname{} is available at \url{https://github.com/dakaidan/CONQuER-Replication}}, a hardware-aware mixed-precision quantisation infrastructure built directly into the compiler pipeline. \toolname{} accepts full-precision \MLIR{} \TOSA{} programs, generates quantised candidates within the compiler, and searches for target-specific mixed-precision policies using a combination of lightweight hardware-aware filtering and direct on-device execution. This framing shifts quantisation from an external, framework-dependent preprocessing step to a unified, compiler-integrated optimisation. By operating natively in this middle layer, \toolname{} provides a single, framework-agnostic entry point that tightly couples candidate generation with backend compiler lowering, ensuring that all explored configurations are physically viable for execution.

The main contributions of this work are:

\begin{enumerate}
    \item \textbf{Native \MLIR{} \TOSA{} Quantisation Infrastructure.} We introduce \texttt{\toolopt{}}, a compiler toolchain that natively generates and manipulates quantised graphs directly at the \TOSA{} dialect level. By operating at this intermediate representation (IR), the toolchain acts as a bridge: it decouples quantisation from restrictive front-end APIs and guarantees that evaluated mixed-precision policies are intrinsically supported by the compiler's downstream lowering passes. This deep integration establishes the necessary foundation for future explorations into quantisation-aware compiler optimisations.

    \item \textbf{Dual-Surrogate Pre-Screening Engine.} To mitigate the prohibitive costs of exhaustive physical profiling, we propose a lightweight filtering mechanism. This engine couples a hardware surrogate (evaluating memory constraints and compute bounds) with an accuracy surrogate (evaluating informational fragility via the feature space isotropy~\cite{leeAZNASAssemblingZeroCost2024} of layer activations) to prune weak configurations prior to compilation and execution.
    
    \item \textbf{Online-Calibrated Hardware-in-the-Loop Search.} We implement an NSGA-II evolutionary search that incorporates an online logarithmic calibrator. By executing only the most promising candidates on physical hardware, the system dynamically aligns the surrogate proxies with ground-truth behaviour across generations.
\end{enumerate}

We evaluate \toolname{} across a diverse spectrum of execution targets, including mobile processors (\targetMobileCPU{}, and \targetLaptopCPU{}), and discrete server hardware (\targetServerGPU{}). Through ablation and comparative analyses, our results demonstrate that effective mixed-precision policies are heavily hardware-dependent. \toolname{} consistently discovers target-specific, Pareto-optimal configurations that yield superior latency, accuracy, and memory trade-offs compared to state-of-the-art hardware-agnostic and hardware-aware baselines. Furthermore, our cross-target transferability studies empirically validate that policies optimised for one hardware profile experience significant degradation when transferred to mismatched targets, underscoring the necessity of our integrated, online-calibrated approach.
\section{\toolname{} Search and Compiler Infrastructure}
\label{sec:tool}


\toolname{} provides a unified compiler infrastructure by shifting mixed-precision quantisation from external front-end scripts directly into the \MLIR{} toolchain. The architecture comprises three primary components: native \MLIR{} \TOSA{} quantisation via \toolopt{}, a dual-surrogate pre-screening engine, and an online-calibrated evolutionary search loop.

\subsection{Native \TOSA{} Quantisation and Profiling}
The core pipeline is driven by \toolopt{}, a custom toolchain that performs quantisation natively at the \MLIR{} \TOSA{} dialect level~\cite{tosa_spec_1_0_1}. By operating on \TOSA{}, the infrastructure remains framework-agnostic and decouples the quantisation process from rigid front-end restrictions. 

Prior to the search phase, \toolopt{} performs an initial calibration and profiling pass. A set of 256 images from the validation set is utilised to compute activation ranges (min/max clipping values) and to extract the covariance matrices of layer activations. The Shannon entropy of the resulting eigenvalues functions as the primary model-aware sensitivity metric, quantifying the isotropy~\cite{leeAZNASAssemblingZeroCost2024} of the layer's feature space to determine its fragility to quantisation noise.

\subsection{Dual-Surrogate Pre-Screening Engine}
To alleviate the computational cost of exhaustive hardware-in-the-loop (HIL) evaluation, \toolname{} employs an over-sampling strategy coupled with a lightweight, dual-surrogate filtering engine. Candidate configurations are evaluated using these proxies before any physical compilation is permitted.

\begin{figure*}[t]
    \centering
    \resizebox{0.8\linewidth}{!}{\begin{tikzpicture}[
        >=stealth,
        font=\sffamily\small,
        hwbox/.style={rectangle, draw=blue!70!black, thick, fill=blue!5, rounded corners=4pt, minimum width=3.2cm, minimum height=0.9cm, align=center},
        IRbox/.style={rectangle, draw=orange!80!black, thick, fill=orange!5, rounded corners=4pt, minimum width=3.2cm, minimum height=0.9cm, align=center},
        calcbox/.style={rectangle, draw=green!60!black, thick, fill=green!5, rounded corners=4pt, minimum width=3.6cm, minimum height=1.1cm, align=center},
        penbox/.style={rectangle, draw=black!60, thick, dashed, fill=gray!5, rounded corners=4pt, minimum width=2.8cm, minimum height=0.8cm, align=center},
        arrow/.style={->, thick, draw=black!80},
        dashedarrow/.style={->, thick, dashed, draw=black!80}
    ]

    \node[hwbox] (hw) at (-3.5, 0) {\textbf{Target Hardware Profile} \\ \textit{\eg{}, A100, i5, Snapdragon}};
    \node[IRbox] (IR) at (3.5, 0) {\textbf{Quantised IR Node} \\ \textit{Op Kind, Shape, Precision}};

    \node[hwbox] (mem_hier) at (-5.8, -1.6) {\textbf{Memory Hierarchy} \\ L1/L2/L3 Caps \& BW};
    \node[hwbox] (comp_dom) at (-1.9, -1.6) {\textbf{Compute Domains} \\ Vector \& Matrix Engines};

    \node[IRbox] (traffic) at (1.9, -1.6) {\textbf{Data Traffic} \\ Working Set Bytes};
    \node[IRbox] (work) at (5.8, -1.6) {\textbf{Compute Workload} \\ Estimated MACs/Ops};

    \draw[arrow, rounded corners=3pt] (hw.south) -- ++(0,-0.3) -| (mem_hier.north);
    \draw[arrow, rounded corners=3pt] (hw.south) -- ++(0,-0.3) -| (comp_dom.north);

    \draw[arrow, rounded corners=3pt] (IR.south) -- ++(0,-0.3) -| (traffic.north);
    \draw[arrow, rounded corners=3pt] (IR.south) -- ++(0,-0.3) -| (work.north);

    \node[calcbox] (mem_lat) at (-4.5, -4.1)
    {\textbf{Memory Latency} ($C_{mem}$) \\
    $\frac{\text{Byte Traffic}}{\text{Effective Bandwidth}}$};

    \node[calcbox] (comp_lat) at (4.5, -4.1)
    {\textbf{Compute Latency} ($C_{comp}$) \\
    $\frac{\text{Compute Ops}}{\text{Target Throughput}}$};

    \draw[arrow] (mem_hier.south) -- (mem_hier.south |- mem_lat.north);
    \draw[arrow] (work.south) -- (work.south |- comp_lat.north);

    \draw[arrow, rounded corners=4pt]
        (comp_dom.south)
        -- ++(0,-0.4)
        -| (comp_lat.north);

    \draw[arrow, rounded corners=4pt, preaction={draw, white, line width=4pt}]
        (traffic.south)
        -- ++(0,-0.2)
        -| (mem_lat.north);

    \node[penbox] (penalties) at (0, -3.2)
    {\textbf{Heuristic Penalties} \\
    Alignment, Layout, \\
    Precision Emulation};

    \draw[dashedarrow, rounded corners=4pt]
        (penalties.west)
        -| ([xshift=1.3cm]mem_lat.north);

    \draw[dashedarrow, rounded corners=4pt]
        (penalties.east)
        -| ([xshift=-1.3cm]comp_lat.north);

    \node[penbox] (dispatch) at (0, -4.4)
    {\textbf{Dispatch Overhead} ($C_{disp}$) \\
    \textit{Device Base Scale}};

    \node[calcbox, minimum width=6.0cm] (total) at (0, -5.8)
    {\textbf{Total Node Latency} \\
    $\max(C_{mem}, C_{comp}) + C_{disp}$};

    \draw[arrow, rounded corners=4pt] (mem_lat.south) |- (total.west);
    \draw[arrow, rounded corners=4pt] (comp_lat.south) |- (total.east);

    \draw[arrow]
        (dispatch.south)
        -- (total.north);

    \node[
        rectangle,
        draw=black!40,
        thick,
        dashed,
        inner sep=12pt,
        rounded corners=6pt,
        fit=(mem_hier) (work) (mem_lat) (comp_lat) (total)
    ] (surrogate) {};

\end{tikzpicture}}
    \caption{Overview of the hardware latency estimation surrogate.}
    \label{fig:hardware_latency_surrogate}
\end{figure*}
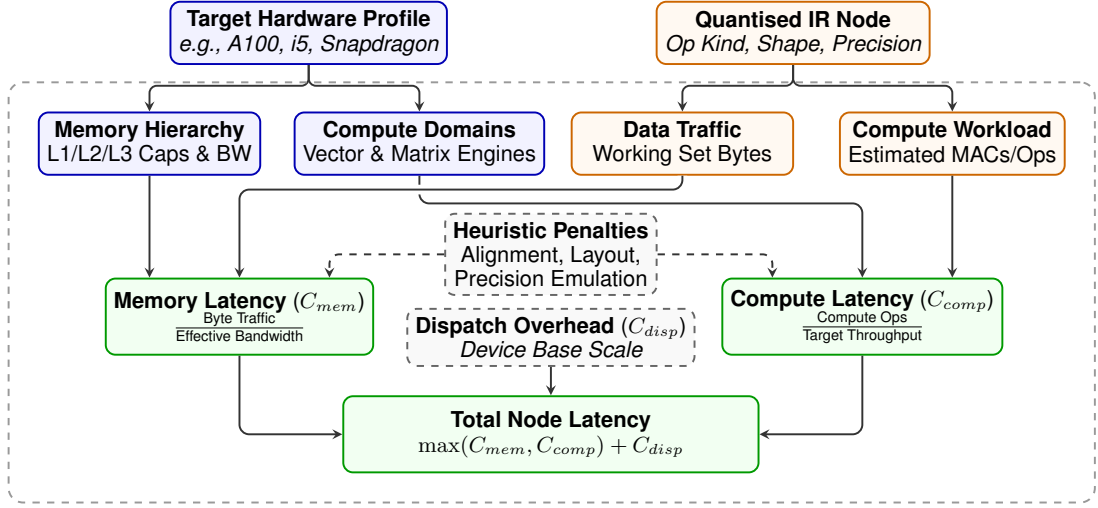

\noindent\textbf{Hardware Surrogate:} For a given candidate policy, \toolname{} generates a Quantised Graph where each node is annotated with precise structural metadata, including operator kind, target precision, tensor dimensions, and memory traffic (activations and constants)~\autoref{fig:hardware_latency_surrogate}. The surrogate estimates theoretical node latency using a roofline model~\cite{williams2009roofline, cabezas2014extending}:
$$C_{total} = \max(C_{compute}, C_{memory}) + C_{dispatch}$$
Compute bounds ($C_{compute}$) are derived from SIMD widths, while memory costs ($C_{memory}$) are mapped against target cache capacities. Configurations that violate the physical memory constraints of the target device are pruned immediately. 

\noindent\textbf{Accuracy Surrogate:} To estimate task accuracy without executing full validation passes, \toolname{} employs an information-theoretic sensitivity proxy inspired by feature space isotropy~\cite{leeAZNASAssemblingZeroCost2024}. For a given layer, the compiler instruments the intermediate tensors to extract the covariance matrix of its activation feature vectors across the channel dimension, $C$. By applying eigenvalue decomposition, we obtain the coefficients of the principal components ($\lambda$). We then derive the normalised Shannon entropy, $\mathcal{H}$, to quantify the isotropy (and therefore the informational fragility) of the layer:
$$\mathcal{H} = -\frac{1}{\ln(C)} \sum_{i=1}^{C} p_i \ln(p_i), \quad \text{where} \quad p_i = \frac{|\lambda_i|}{\sum |\lambda_j|}$$
A higher entropy ($\mathcal{H} \to 1$) indicates an isotropic feature space where all principal components are equally critical, meaning the layer is highly fragile to perturbations. Conversely, a lower entropy implies dimensional redundancy (where the feature space is dominated by a few principal components) and greater robustness. Drawing on a first-order Taylor approximation, the total accuracy degradation penalty for a node is calculated as the sum of its activation and weight penalties. Each penalty is formulated as the product of this entropy-based fragility and the theoretical quantisation noise amplitude, approximated as $\approx 2^{-b}$ for a $b$-bit integer format (and proportionally scaled to account for formats with lower dynamic range).

\subsection{Online-Calibrated Evolutionary Search}
The allocation of mixed-precision configurations is formulated as a multi-objective optimisation problem (minimising latency, accuracy degradation) and navigated using the NSGA-II algorithm~\cite{deb2002fast}. 
As static proxies are susceptible to runtime anomalies and backend-specific compiler optimisations, \toolname{} incorporates an online calibration mechanism. During each generation, the most promising candidates identified by the surrogate engine are fully compiled and executed on the target hardware via \IREE{}, using a distinct 512-image search split. Accuracy is measured directly from these hardware executions, while latency is estimated from a single input using $3$ warm-up iterations followed by $10$ timed runs. The resulting measurements are then incorporated into an online logarithmic calibrator, $y = a \ln(x) + b$, which uses a sliding window of recent evaluations to continuously align the surrogate models with observed behaviour across successive generations.
\section{Evaluation}
\label{sec:evaluation}

The evaluation of \toolname{} is structured to assess its optimisation efficacy, search efficiency, component contributions, and hardware transferability. Specifically, the experiments address the following research questions:

\begin{itemize}
    \item \textbf{RQ1 (Comparative Efficacy):} To what extent does \toolname{} improve the Pareto-optimal trade-offs (latency, accuracy, and memory footprint) compared to state-of-the-art hardware-aware and hardware-agnostic quantisation methods across varied target devices?
    \item \textbf{RQ2 (Cross-Target Transferability):} Do hardware-specific quantisation policies generated by \toolname{} demonstrate significant performance degradation when deployed on mismatched hardware targets, thereby necessitating target-specific search?
    \item \textbf{RQ3 (Ablation Analysis):} What are the individual and combined contributions of the hardware surrogate and the accuracy surrogate to the quality of the discovered configurations and overall search efficiency?
\end{itemize} 

\subsection{Experimental Setup}

\textbf{Datasets and Profiling Isolation:} Experiments are conducted using the ImageNet dataset. To prevent quantisation overfitting, strict data isolation is enforced. A 256-image subset is reserved exclusively for the initial \texttt{\toolopt{}} calibration and sensitivity profiling. A separate, 512-image subset is utilised for the HIL evaluation within the genetic algorithm. Final Pareto-optimal configurations are independently tested on the remaining unseen images from the validation set.

\textbf{Target Hardware and Benchmarks:} 
The evaluation spans multiple target architectures to accurately reflect diverse deployment environments. The model benchmarks include MobileNetV2, ResNet-18, ResNet-50, and efficientnet.
The experimental executions are partitioned as follows:
\begin{itemize}
    \item \textit{Full \toolname{} Execution (RQ1, RQ2):} Executed on an \targetLaptopCPU{} and a \targetMobileCPU{}~\cite{qualcomm_snapdragon_8_elite} using MobileNetV2, ResNet-18, and ResNet-50 to represent edge deployment constraints.
    \item \textit{Ablation Studies (RQ3):} Executed on a discrete Server GPU (\targetServerGPU{}) using the full suite of benchmark models.
\end{itemize}

\textbf{Baselines and SOTA Comparison:}
Configurations discovered by \toolname{} are benchmarked against three primary baselines across all applicable hardware targets:
\begin{enumerate}
    \item \textit{FP32 Baseline:} The unquantised \MLIR{} execution, which serves as the upper bound for model accuracy and the baseline for latency and memory footprint.
    \item \textit{InfoQ \cite{akbulutInfoQMixedPrecisionQuantization2026}:} An analytical approach that measures global information flow and uses Integer Linear Programming (ILP) to allocate bit-widths under a specified resource budget. Due to runtime incompatibilities on ARM architectures, InfoQ's execution latency is evaluated natively using the PyTorch runtime rather than the compiled \IREE{} stack, and results are restricted to the \targetServerGPU{} and \targetLaptopCPU{}.
    \item \textit{SeQTO \cite{louloudakisSelectiveQuantizationTuner2025}:} A direct Hardware-in-the-Loop (HIL) method that uses a greedy, layer-by-layer tuning strategy. To ensure a fair comparison, the physical evaluation budget for both SeQTO and \toolname{} was strictly capped at a maximum of 1,440 hardware inferences.
\end{enumerate}

\textbf{Methodology for Transferability and Ablation:}
To address RQ2, optimal policies derived for a specific hardware target (e.g. \targetServerGPU{}) are compiled and executed on mismatched target (e.g., \targetLaptopCPU{}) to quantify latency and memory penalties incurred by hardware-agnostic transfer.

To address RQ3, the search infrastructure is evaluated under four surrogate configurations: No Proxy (pure HIL), Accuracy Proxy Only, Hardware Proxy Only, and Dual Proxy (Both). 

Due to the stochastic nature of evolutionary algorithms, all executions for \toolname{}, InfoQ, SeQTO, and the ablation variants are repeated 5 times with varying random seeds. Results report the mean and variance.
\section{Results}
\label{sec:results}

\subsection{RQ1: Comparative Efficacy}
\label{sec:rq1_comparative}

Below we benchmark \toolname{} against two SOTA approaches: SeQTO, and InfoQ. Each approach is evaluated across three hardware targets and models to evaluate latency and accuracy trade-offs.

\vspace{1em}
\noindent\textbf{Benchmark Comparison: SeQTO}

\begin{table*}[t]
\centering
\footnotesize
\setlength{\tabcolsep}{4pt}
\caption{\toolname{} vs. SeQTO. Speedups are relative to the unquantised FP32 baseline. The `Best' columns indicate the fastest configuration maintaining $\le 5\%$ accuracy degradation (falling back to $\le 15\%$ if a framework fails to meet the threshold). Averages include 5 reps ($\pm$ SD).}
\label{tab:RQ1:SeQTO}
\begin{tabular}{ll | cc cc | cc cc}
\toprule
\multirow{2}{*}{\textbf{Target}} & \multirow{2}{*}{\textbf{Model}} & \multicolumn{4}{c|}{\textbf{\toolname{} (Ours)}} & \multicolumn{4}{c}{\textbf{SeQTO [3]}} \\
\cmidrule(lr){3-6} \cmidrule(lr){7-10}
 & & \multicolumn{2}{c}{Best Case} & \multicolumn{2}{c|}{Average ($\pm$ SD)} & \multicolumn{2}{c}{Best Case} & \multicolumn{2}{c}{Average ($\pm$ SD)} \\
 & & Spd. & Acc $\downarrow$ & Spd. & Acc $\downarrow$ & Spd. & Acc $\downarrow$ & Spd. & Acc $\downarrow$ \\
\midrule
\textbf{\targetServerGPU{}}         & MobileNetV2  & 2.31$\times$ &     2.08\% & 2.05$\pm$0.16$\times$ &   2.23$\pm$11.12\% & 1.33$\times$ &    10.82\% & 1.31$\pm$0.02$\times$ &   10.38$\pm$0.44\% \\
                             & ResNet-18    & 1.57$\times$ &     0.46\% & 1.38$\pm$0.21$\times$ &  4.34$\pm$3.62\% & 1.17$\times$ &     3.86\% &         1.16$\pm$0.18$\times$ &    3.97$\pm$0.27\% \\
                             & ResNet-50    & 12.19$\times$ &     1.44\% & 7.52$\pm$3.82$\times$ &   6.57$\pm$1.29\% & 1.49$\times$ &     5.38\% & 1.44$\pm$0.06$\times$ &    5.41$\pm$0.05\% \\
\midrule
\textbf{\targetLaptopCPU{}}       & MobileNetV2  & \textcolor{red}{0.96$\times$} &     0.39\% & 0.83$\pm$0.05$\times$ &    4.04$\pm$3.58\% & \textcolor{red}{0.95$\times$} &     0.00\% & 0.92$\pm$0.04$\times$ &     1.42$\pm$1.2\% \\
                             & ResNet-18    & \textcolor{red}{0.98$\times$} &     1.56\% & 0.80$\pm$0.08$\times$ &   4.91$\pm$2.88\% & \textcolor{red}{0.95$\times$} &     0.00\% & 0.90$\pm$0.07$\times$ &      0.99$\pm$0.45\% \\
                             & ResNet-50    & \textcolor{red}{0.84$\times$} &     0.00\% & $0.72\pm0.04\times$ &   2.47$\pm$1.79\% & \textcolor{red}{0.95$\times$} &     0.00\% & 0.92$\pm$0.03$\times$ &       2.90$\pm$1.93\% \\
\midrule
\textbf{\targetMobileCPU{}}  & MobileNetV2  & 1.25$\times$ &     0.78\% & 0.93$\pm$0.12$\times$ &    0.23$\pm$0.32\% & \textcolor{red}{0.93$\times$} &     6.10\% & 0.69$\pm$0.29$\times$ &    8.82$\pm$2.14\% \\
                             & ResNet-18    & 3.31$\times$ &     0.00\% & 2.61$\pm$0.62$\times$ &  1.77$\pm$1.04\% & 1.82$\times$ &     0.00\% & 1.09$\pm$0.56$\times$ &    1.59$\pm$1.87\% \\
                             & ResNet-50    & 1.51$\times$ &     0.00\% & 1.21$\pm$0.29$\times$ &  4.13$\pm$0.71\% & 3.37$\times$ &     3.58\% & 3.35$\pm$0.02$\times$ &    3.99$\pm$0.29\% \\
\midrule
\bottomrule
\end{tabular}
\end{table*}

We benchmark \toolname{} against SeQTO, a greedy, layer-by-layer hardware-in-the-loop (HIL) tuning method. Because SeQTO is deterministic and generates a single configuration per run, both approaches were evaluated across 5 repetitions using different random seeds to account for execution variance. Table~\ref{tab:RQ1:SeQTO} details their performance across three hardware targets. The speedup is relative to the unquantised (PyTorch to MLIR) FP32 baseline and the corresponding drop in top-1 accuracy. ``Best Case'' configurations denote the fastest policy maintaining $\le 5\%$ accuracy degradation, falling back to $\le 15\%$ if a framework cannot find a compliant configuration. 

While SeQTO outputs standard ONNX models, we observed that certain configurations generated by SeQTO structurally failed to lower through the \IREE{} compiler pipeline. This highlights a limitation of decoupled optimisation, and the fragmented nature of model quantisation at the framework level: without integrating the target compiler directly into the search loop, theoretically valid MPQ policies can produce malformed or unsupported graphs that silently fail at deployment time.

Across the configurations, all solutions which, when evaluated, resulted in accuracy loss over 15\% were excluded as outliers. No such cases were present with SeQTO, which always found reasonable solutions. Across all reps, there were some outlier cases present with \toolname{}; however, these were limited to a minority of the Pareto front, representing no more than 1 of the resulting optimal results per rep.

\textbf{Performance on Discrete GPU (\targetServerGPU{})} 
On the \targetServerGPU{}, \toolname{} outperforms SeQTO in both latency reduction and accuracy preservation. This is most pronounced on ResNet-50, where \toolname{} achieves a $12.19\times$ speedup with a $1.44\%$ accuracy drop, whereas SeQTO reaches a $1.49\times$ speedup with a $5.38\%$ drop. On MobileNetV2, SeQTO fails to meet the $5\%$ accuracy threshold, falling back to a policy with a $10.82\%$ accuracy drop and a $1.33\times$ speedup. \toolname{} identifies a compliant policy with a $2.31\times$ speedup and a $2.08\%$ accuracy drop.

\textbf{Performance on Mobile CPU (\targetMobileCPU{})} 
On the \targetMobileCPU{}, \toolname{} finds a policy for ResNet-18 that yields a $3.31\times$ speedup with $0.00\%$ accuracy degradation, compared to SeQTO's $1.82\times$ speedup. For MobileNetV2, SeQTO again misses the $5\%$ bound, falling back to a configuration with a $6.10\%$ accuracy drop and a latency regression ($0.93\times$ speedup). \toolname{} finds a configuration with a $1.25\times$ speedup and a $0.78\%$ accuracy drop. On ResNet-50, SeQTO achieves a higher peak speedup ($3.37\times$ vs. \toolname{}'s $1.51\times$) but incurs a $3.58\%$ accuracy penalty, whereas \toolname{} maintains a $0.00\%$ accuracy loss.

\textbf{Performance on Laptop CPU (\targetLaptopCPU{})} 
The \targetLaptopCPU{} environment introduces a fundamental hardware constraint: the runtime unpacking penalty for storing parameters in FP16, which requires upcasting to 32-bit~\cite{intel_isa_extensions}. Consequently, neither \toolname{} nor SeQTO achieve an absolute speedup ($\ge 1.0\times$) over the FP32 baseline while maintaining high accuracy. However, \toolname{} manages this latency regression more effectively on compact models. For MobileNetV2 and ResNet-18, \toolname{} limits the slowdown to $0.96\times$ and $0.98\times$, trading marginal accuracy drops ($0.39\%$ and $1.56\%$) to reduce the unpacking penalty. SeQTO maintains a $0.00\%$ accuracy loss but exhibits a uniform $0.95\times$ slowdown across all three models.
These results demonstrate that SeQTO's greedy, layer-by-layer methodology frequently traps the search in suboptimal local minima, resulting in either greater accuracy degradation or restricted latency improvements. By employing a globally aware search that strictly bounds accuracy loss while prioritising latency, \toolname{} successfully circumvents these local minima to discover superior, hardware-specific optima across diverse architectures and model scales.

\vspace{1em}
\noindent\textbf{Benchmark Comparison: InfoQ}

\begin{table*}[t]
\centering
\footnotesize
\setlength{\tabcolsep}{4pt}
\caption{CoNQuER vs. InfoQ. Speedups are relative to the unquantised FP32 baseline. The `Best' columns indicate the fastest configuration maintaining $\le 5\%$ accuracy degradation. Absolute latency is reported alongside relative speedup.}
\label{tab:RQ1:InfoQ}
\begin{tabular}{ll | ccc | ccc}
\toprule
\multirow{2}{*}{\textbf{Method}} & \multirow{2}{*}{\textbf{Metric}} & \multicolumn{3}{c|}{\textbf{NVIDIA A100 80GB}} & \multicolumn{3}{c}{\textbf{Intel I5 1145g7}} \\
\cmidrule(lr){3-5} \cmidrule(lr){6-8}
 & & MobileNetV2 & ResNet-18 & ResNet-50 & MobileNetV2 & ResNet-18 & ResNet-50 \\
\midrule
\multirow{3}{*}{\textbf{CoNQuER (Ours)}} 
 & Latency & 6.78 ms & 8.42 ms & 8.43 ms & 77.17 ms & 53.69 ms & 69.06 \\
 & Speedup & 2.31$\times$ & 1.57$\times$ & 12.19$\times$ & \textcolor{red}{0.96$\times$} & \textcolor{red}{0.98$\times$} & \textcolor{red}{0.84$\times$} \\
 & Acc $\downarrow$ & 2.08\% & 0.46\% & 1.44\% & 0.39\% & 1.56\% & 0.00\% \\
\midrule
\multirow{4}{*}{\textbf{InfoQ (4x)}} 
 & Latency & 69.08 ms & 20.74 ms & 52.18 ms & 88.72 ms & 62.15 ms & 60.77 ms \\
 & Speedup & \textcolor{red}{0.48$\times$} & \textcolor{red}{0.96$\times$} & \textcolor{red}{0.77$\times$} & \textcolor{red}{0.96$\times$} & 1.14$\times$ & 1.39$\times$ \\
 & Acc $\downarrow$ & 21.12\% & 5.58\% & 4.34\% & 21.66\% & 6.02\% & 4.24\% \\
 & QAT Acc $\downarrow$ & +0.96\% & +0.04\% & -0.32\% & +0.96\% & +0.12\% & -0.32\% \\
\midrule
\multirow{4}{*}{\textbf{InfoQ (6x)}} 
 & Latency & 47.59 ms & 20.55 ms & 54.00 ms & 174.88 ms & 69.45 ms & 50.05 ms \\
 & Speedup & \textcolor{red}{0.70$\times$} & \textcolor{red}{0.97$\times$} & \textcolor{red}{0.74$\times$} & \textcolor{red}{0.49$\times$} & 1.02$\times$ & 1.68$\times$ \\
 & Acc $\downarrow$ & 99.92\% & 8.72\% & 92.46\% & 99.90\% & 8.90\% & 92.38\% \\
 & QAT Acc $\downarrow$ & -0.82\% & +1.22\% & -0.02\% & -0.82\% & +1.24\% & +0.12\% \\
\midrule
\multirow{4}{*}{\textbf{InfoQ (high)}} 
 & Latency & 48.23 ms & 20.90 ms & 54.89 ms & 482.51 ms & 36.36 ms & 52.86 ms \\
 & Speedup & \textcolor{red}{0.69$\times$} & \textcolor{red}{0.95$\times$} & \textcolor{red}{0.73$\times$} & \textcolor{red}{0.18$\times$} & 1.95$\times$ & 1.59$\times$ \\
 & Acc $\downarrow$ & 99.92\% & 99.84\% & 99.88\% & 99.92\% & 99.86\% & 99.88\% \\
 & QAT Acc $\downarrow$ & -0.90\% & -0.02\% & -1.04\% & -0.86\% & -0.02\% & -1.04\% \\
\midrule
\bottomrule
\end{tabular}
\end{table*}

We evaluate InfoQ as a representative hardware-agnostic analytical mixed-precision framework. InfoQ computes mixed-precision allocations analytically in a matter of minutes, whereas \toolname{}'s hardware-in-the-loop evolutionary search requires several hours of compilation and physical execution. However, this upfront speed comes with severe deployment penalties that hinder practical use without Quantisation-Aware Training (QAT).
Unlike \toolname{}, InfoQ outputs PyTorch \texttt{ScriptModule} artifacts that rely on legacy quantisation and backend-specific operator containers. We observed that these artifacts structurally fail to lower through modern Ahead-Of-Time (AOT) edge compilers, including \IREE{} (\texttt{torch-mlir}) and ExecuTorch~\cite{executorch2026} (\texttt{torch.export}). This translation failure is driven by InfoQ's utilisation of sub-byte bit-widths (e.g., 2-bit, 4-bit) that lack corresponding MLIR legalisations, as well as its reliance on \texttt{fbgemm} operators that are strictly bound to x86 instruction sets. Consequently, deploying InfoQ policies to the ARM-based \targetMobileCPU{} is completely unsupported.

To explore the latency-accuracy trade-offs of this analytical approach, we evaluate InfoQ across three compression targets: 4$\times$, 6$\times$, and `high' (maximum logical compression, exceeding 10$\times$ depending on model scale). Within InfoQ's analytical formulation, these targets act as strict memory constraints applied to its Integer Linear Programming (ILP) solver. By setting these bounds, the ILP solver is forced to allocate mixed-precision bit-widths such that the resulting model's memory footprint is guaranteed to be at least 4 times, 6 times, or maximally smaller than the unquantised 32-bit baseline.

We first evaluate these generated policies under strict Post-Training Quantisation (PTQ) constraints to directly compare against \toolname{}'s zero-retraining paradigm. However, because analytical methods like InfoQ aggressively truncate precision without hardware-in-the-loop feedback, they typically rely on subsequent Quantisation-Aware Training (QAT) to recover functional accuracy. To evaluate this recovery, we also report InfoQ's accuracy after applying QAT on the \targetServerGPU{}. We restrict this QAT phase to a 12-hour training budget to match the average wall-clock execution time of \toolname{}'s evolutionary search. It is important to note that this is not a perfectly equivalent resource comparison: completing \toolname{}'s search within 12 hours primarily relies on sufficient standard CPU compute to concurrently compile candidates, while executing 12 hours of QAT requires continuous, access to dedicated training hardware.

The results in Table~\ref{tab:RQ1:InfoQ} demonstrate that, without hardware-aware compiler integration, InfoQ frequently yields policies that struggle to balance latency improvements with accuracy retention in a PTQ setting. On the \targetServerGPU{}, InfoQ universally fails to achieve an absolute speedup ($\ge 1.0\times$) over the FP32 baseline across all models and compression levels. For example, on ResNet-50, InfoQ's 4$\times$ configuration operates at a $0.77\times$ speedup (a latency regression) with a $4.34\%$ accuracy drop, while \toolname{} achieves a $12.19\times$ speedup with only a $1.44\%$ drop. On MobileNetV2, InfoQ's 4$\times$ policy incurs a $21.12\%$ accuracy penalty and a $0.48\times$ slowdown, compared to \toolname{}'s $2.31\times$ speedup and $2.08\%$ drop.

On the \targetLaptopCPU{}, while InfoQ manages marginal speedups on ResNet architectures, its aggressive use of sub-byte truncation naturally leads to significant accuracy degradation under a zero-retraining constraint. For instance, the 6$\times$ and `high' compressions on MobileNetV2 exhibit $99.90\%$ and $99.92\%$ degradation, respectively. This expected drop at high compression rates demonstrates a core limitation of the framework: these analytical policies require subsequent QAT to become viable, pushing the computational burden further down the pipeline. However, when QAT is applied, any regression in accuracy is resolved, and often the resultant model achieves higher accuracy on the training dataset than the original model.
Ultimately, while InfoQ's analytical approach is remarkably fast, these benchmarks highlight a critical software engineering trade-off. Investing computational budget into a compiler-integrated, hardware-in-the-loop search guarantees the generation of executable, high-performance policies that natively preserve accuracy, whereas rapid hardware-agnostic methods consistently yield artefacts that are not always fully compatible or that necessitate additional resource-intensive phases like QAT to recover functional performance. 

\subsection{RQ2: Cross-Target Transferability}
\label{sec:rq3_transferability}

\begin{figure*}[t]
    \centering
    \includegraphics[width=\linewidth]{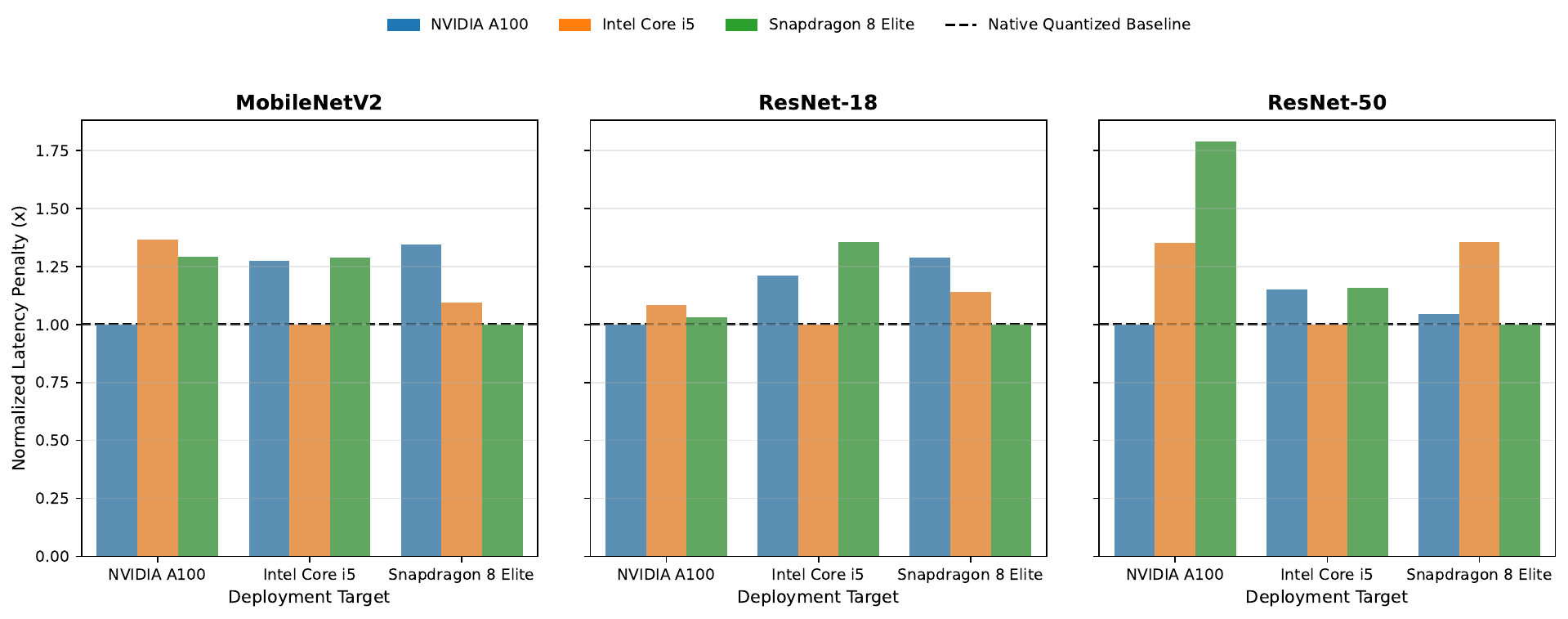}
    \caption{Cross-target transferability of mixed-precision policies across \targetServerGPU{}, \targetLaptopCPU{}, and \targetMobileCPU{}. The y-axis shows the normalised latency penalty incurred when deploying a policy optimised for the source hardware onto a mismatched target. The black dashed line (1.0$\times$) represents the native hardware-aware optimal policy, while the red dotted line indicates the unquantised FP32 baseline.}
    \label{fig:RQ2:Transfer}
\end{figure*}

\begin{table}[!hbt]
\caption{FP32 Baseline Latency Multiples relative to Native Quantised Optimums}
\label{tab:fp32_multiples}
\begin{tabular}{lrrr}
\toprule
 & MobileNetV2 & ResNet-18 & ResNet-50 \\
\midrule
\targetLaptopCPU{} & 0.74 & 0.91 & 0.74 \\
\targetServerGPU{} & 2.20 & 1.53 & 13.44 \\
\targetMobileCPU{} & 1.02 & 2.99 & 1.45 \\
\bottomrule
\end{tabular}
\end{table}

To evaluate cross-target transferability, we extracted three representative mixed-precision policies from the Pareto front of each source hardware and model combination: the latency-optimal policy, the accuracy-optimal policy, and a balanced middle-ground policy. These policies were deployed onto mismatched target architectures, and their resulting relative latencies were averaged to quantify the transfer penalty. Crucially, across all evaluations, the best latency achieved by a transferred non-native policy never surpasses the best latency of the natively optimised policy. While a heavily quantised non-native policy might occasionally execute faster than a native policy optimised strictly for accuracy, when evaluating equivalent Pareto objectives, the hardware-aware native search remains strictly superior.

Figure~\ref{fig:RQ2:Transfer} reports the normalised latency penalties incurred by these hardware-agnostic transfers, using the natively optimised policy's latency as the $1.0\times$ baseline. This demonstrates that deploying a policy on a non-native target consistently results in a latency penalty across all evaluated models and hardware combinations.

On the \targetServerGPU{}, executing a MobileNetV2 policy optimised for the \targetLaptopCPU{} incurs a $1.37\times$ latency penalty, while transferring a \targetMobileCPU{}-optimised policy yields a $1.29\times$ penalty. This regression is particularly pronounced on ResNet-50, where deploying the \targetMobileCPU{} policy on the \targetServerGPU{} results in a $1.79\times$ increase in latency over the native baseline. Conversely, transferring an \targetServerGPU{}-optimised policy to the \targetMobileCPU{} introduces penalties ranging from $1.05\times$ (ResNet-50) to $1.34\times$ (MobileNetV2). These variations indicate that quantisation policies tightly overfit to the specific compute and memory hierarchies of their target architecture.

The relative severity of the transfer penalty is exacerbated by model scale and baseline execution time. For highly parametrised models such as ResNet-50, the unquantised FP32 latency is inherently high (e.g., 104.97~ms on the \targetServerGPU{}), meaning any valid mixed-precision configuration provides a substantial absolute speedup (reducing to 24.11~ms natively). Because the baseline latency is large, the relative differences between a natively optimal policy and a transferred policy are compressed, yielding smaller relative penalties (such as the $1.05\times$ penalty when transferring from the \targetServerGPU{} to the \targetMobileCPU{}). Nonetheless, the native search consistently identifies the most performant execution path.

An exception to the general latency improvement over FP32 is observed on the \targetLaptopCPU{} target. As shown in Figure~\ref{fig:RQ2:Transfer}, all quantised policies on this hardware execute slower than the unquantised FP32 baseline (indicated by the red dotted line). Maintaining accuracy on this model often involves the retention of float types (e.g., FP16) in many of the solutions. However, FP16 is purely a storage type on this processor and must be unpacked to 32-bit at runtime, incurring an execution overhead that exceeds native FP32 operations.

Despite these hardware constraints, target-specific search continues to provide measurable benefits over agnostic transfer. While the native search cannot overcome the fundamental unpacking penalty to beat the FP32 baseline, it explicitly minimises the resulting latency regression while reducing the model size. On MobileNetV2, the native i5 policy executes in 40.26~ms. In contrast, transferring policies optimised for the \targetServerGPU{} or \targetMobileCPU{} -- which are structurally unaware of this unpacking penalty -- increases execution time to 50.00~ms and 51.76~ms, respectively. This results in relative penalties of $1.27\times$ and $1.29\times$ over the native quantised configuration. This demonstrates that direct hardware-in-the-loop optimisation is necessary to navigate target-specific bottlenecks, whether to maximise performance gains or to minimise architectural regressions.

\subsection{RQ3: Ablation Analysis}
\label{sec:rq2_ablation}

\begin{figure*}[t]
\centering
\includegraphics[width=0.75\textwidth]{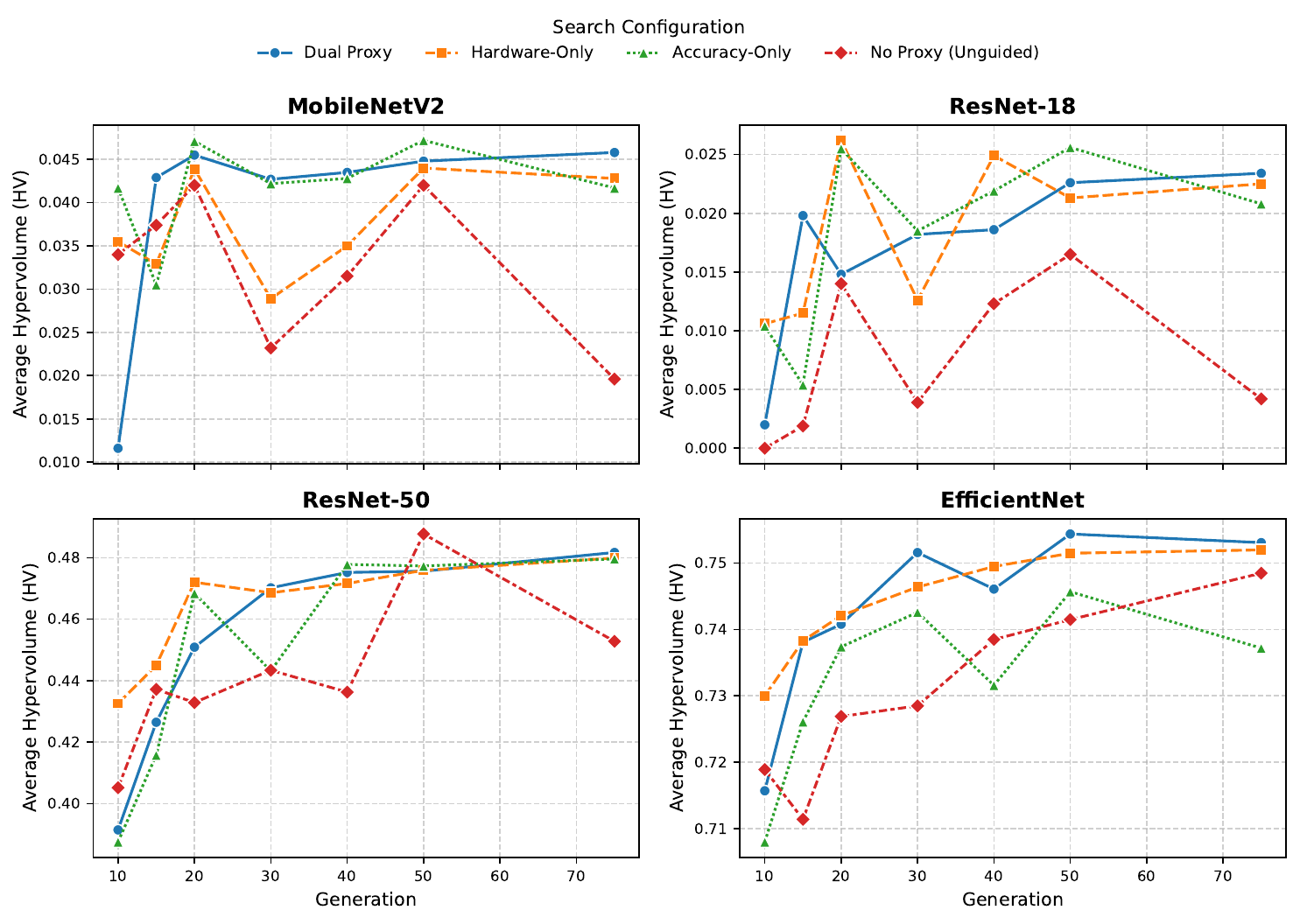}
\caption{Generational trajectory of the average hypervolumes (normalised) vs. Generation across ablation configurations. Single proxies and unguided searches exhibit higher instability, whereas the dual-surrogate approach achieves monotonic convergence.}
\label{fig:RQ3:Trajectories}
\end{figure*}

To understand the individual and combined impact of \toolname{}'s pre-screening mechanisms, we ablate the surrogate engine into four configurations: unguided search (\textit{No Proxy}), accuracy-guided only (\textit{Accuracy-Only}), hardware-latency-guided only (\textit{Latency-Only}), and the complete dual-surrogate system (\textit{Dual}). The ablation was executed on an \targetServerGPU{} across all benchmark models. Table~\ref{tab:RQ3:ablation_metrics} summarises the convergence speed, Hypervolume (HV), and Pareto front size, while Figure~\ref{fig:RQ3:Trajectories} illustrates the generational search stability.

\textbf{Unguided Search:} 
Naive evolutionary search relying solely on HIL evaluations (\textit{No Proxy}) proves highly inefficient for MPQ. Without analytical surrogates to pre-filter the combinatorial search space, the genetic algorithm expends its evaluation budget on fundamentally non-viable configurations. As shown in Table~\ref{tab:RQ3:ablation_metrics}, this lack of guidance leads to search instability. On ResNet-18, the unguided search exhibits a high convergence variance ($\pm$ 37.0 generations) and yields a heavily degraded average HV of 0.0042, compared to 0.0234 for the dual-proxy approach. This confirms that physical HIL feedback alone is too sparse and noisy to effectively navigate the MPQ space within practical compilation budgets.

\textbf{Single Proxies:} 
Introducing a single surrogate substantially improves upon the unguided baseline, but it exposes the search to objective skew (surrogate overfitting). When the algorithm is constrained along only one dimension, the search naturally biases away from the unconstrained dimension. While the HIL evaluation partially corrects this during the evolutionary loop, the overall generational trajectory remains skewed.

When relying exclusively on the accuracy surrogate (\textit{Accuracy-Only}), the search generates populations with minimal accuracy degradation. However, because the surrogate is blind to physical execution bottlenecks, it frequently selects high-precision configurations that map poorly to the target hardware, resulting in median latencies up to $1.24\times$ slower than the optimal front on architectures like ResNet-50. Conversely, the hardware-only surrogate (\textit{Hardware-latency-Only}) aggressively quantises to maximise arithmetic throughput; this achieves minimal latency but incurs notable accuracy drops, resulting in $3.1\times$ worse accuracy loss from the baseline when compared to the dual-proxy. 

Consequently, single-proxy configurations skew the Pareto front heavily toward their respective objectives. They struggle to populate the ``knee" of the curve where balanced trade-offs exist, and their search trajectories are often erratic. They may occasionally discover high-HV configurations but struggle to consistently find them. However, both single-proxy approaches still approach the unconstrained search performance. In deployment scenarios where either latency or accuracy is strictly prioritised over the other, single-proxy guidance remains a viable strategy for finding edge solutions.

\textbf{Dual-Surrogate:} 
The complete dual-surrogate mitigates this objective skew. By bounding the search space with theoretical memory constraints and informational fragility, the system establishes a constrained, viable search space. This prevents generational regression and enforces a smooth, monotonically improving search trajectory (Figure~\ref{fig:RQ3:Trajectories}) that is robust to HIL measurement noise. The dual-surrogate configuration sustains productive exploration, averaging convergence at generation $34.45$ across all evaluated models to consistently yield the highest-quality, most balanced Pareto fronts.

\textbf{Hardware Proportionality and Dispatch Overhead:} 
An important behaviour emerged during the \targetServerGPU{} ablation. Because the \targetServerGPU{} is a highly parallel discrete GPU designed for massive workloads, the execution time of compact models (such as MobileNetV2 and ResNet-18) is heavily bounded by kernel dispatch overhead rather than arithmetic compute. This dispatch overhead introduces runtime noise that can obscure the actual computation time. Consequently, the absolute latency gains from mixed-precision quantisation are compressed on these smaller models compared to larger architectures like ResNet-50. This highlights a crucial deployment consideration: mixed-precision latency gains are most pronounced when the model's computational intensity is proportional to the target hardware's compute capacity -- when this is not the case, other approaches may be superior.

\begin{table}[htbp]
\centering
\caption{Ablation of surrogate components on convergence, Hypervolume (HV), and Pareto front size. Convergence Generation is defined as the generation where the Pareto front reaches within 5\% of its final state. Results show mean $\pm$ standard deviation for 5 repetitions.}
\label{tab:RQ3:ablation_metrics}
\renewcommand{\arraystretch}{1.1}
\resizebox{\columnwidth}{!}{%
\begin{tabular}{@{}lccc@{}}
\toprule
\textbf{Configuration} & \textbf{Convergence Gen.} & \textbf{Hyper-Volume} & \textbf{Front Size} \\ 
\midrule
\multicolumn{4}{c}{\textbf{MobileNetV2}} \\
\midrule
No Proxy & $33.0 \pm 32.2$ & $0.0196 \pm 0.0202$ & $3.2 \pm 1.3$ \\
Accuracy-Only & $45.8 \pm 18.8$ & $0.0417 \pm 0.0022$ & $1.8 \pm 0.4$ \\
Latency-Only & $50.6 \pm 10.7$ & $0.0428 \pm 0.0013$ & $2.2 \pm 1.1$ \\
\textbf{Dual (\toolname{})} & $\mathbf{32.4 \pm 12.8}$ & $\mathbf{0.0458 \pm 0.0015}$ & $\mathbf{3.0 \pm 0.7}$ \\ 
\midrule
\multicolumn{4}{c}{\textbf{ResNet-18}} \\
\midrule
No Proxy & $43.4 \pm 27.0$ & $0.0042 \pm 0.0058$ & $4.2 \pm 1.9$ \\
Accuracy-Only & $51.4 \pm 29.7$ & $0.0208 \pm 0.0024$ & $3.6 \pm 0.9$ \\
Latency-Only & $55.8 \pm 8.3$ & $0.0225 \pm 0.0012$ & $4.4 \pm 1.1$ \\
\textbf{Dual (\toolname{})} & $\mathbf{40.2 \pm 10.5}$ & $\mathbf{0.0234 \pm 0.015}$ & $\mathbf{6.0 \pm 1.0}$ \\ 
\midrule
\multicolumn{4}{c}{\textbf{ResNet-50}} \\
\midrule
No Proxy & $42.6 \pm 29.6$ & $0.4528 \pm 0.0323$ & $7.1 \pm 2.3$ \\
Accuracy-Only & $36.0 \pm 26.1$ & $0.4796 \pm 0.0048$ & $7.4 \pm 1.6$ \\
Latency-Only & $32.8 \pm 20.9$ & $0.4799 \pm 0.0072$ & $7.2 \pm 1.9$ \\
\textbf{Dual (\toolname{})} & $\mathbf{27.4 \pm 13.5}$ & $\mathbf{0.4817 \pm 0.0093}$ & $\mathbf{7.8 \pm 1.3}$ \\ 
\midrule
\multicolumn{4}{c}{\textbf{EfficientNet}} \\
\midrule
No Proxy & $41.8 \pm 18.1$ & $0.7485 \pm 0.0071$ & $7.2 \pm 1.5$ \\
Accuracy-Only & $38.0 \pm 16.2$ & $0.7372 \pm 0.0209$ & $6.0 \pm 1.6$ \\
Latency-Only & $39.8 \pm 15.3$ & $0.7520 \pm 0.0083$ & $6.0 \pm 1.6$ \\
\textbf{Dual (\toolname{})} & $\mathbf{37.8 \pm 13.7}$ & $\mathbf{0.7531 \pm 0.0113}$ & $\mathbf{7.8 \pm 1.6}$ \\ 
\midrule
\end{tabular}%
}
\end{table}

\section{Threats to Validity}
\label{sec:threatstovalidity}

\subsection{Internal Validity} 

\noindent\textbf{Toolchain and runtime mismatches:} Our baselines rely on different frontend pipelines (\eg, PyTorch to \IREE, or \ONNX to \IREE). This results in subtly different versions of the model in \MLIR; which may result in different performance profiles. This reflects the severely fragmented nature of modern ML deployment toolchains. \toolname deliberately shifts quantisation directly into the compiler pipeline at the \TOSA level to avoid these brittle conversions and to ensure optimisation is performed on the relevant deployment artefact. In cases where baselines fail to export or map optimally to \IREE runtimes, this represents the practical limitations developers face. 

\noindent\textbf{Prototype limitations:} \toolopt is a prototype; any suboptimal lowering of specific quantised nodes might artificially penalise those configurations during search. However, this establishes a strict lower bound on performance -- future compiler optimisations may improve our reported gains. 

\noindent\textbf{Thermal management:} Running intensive hardware-in-the-loop (HIL) evaluations on edge devices risks thermal throttling. We mitigated this by using fixed performance profiles on the \targetLaptopCPU and \targetMobileCPU. Further, in the case of \targetServerGPU, the models being run are far below the capacity of such a GPU, and as such, we never approached thermal limits in any of the runs.

\subsection{External Validity} 

\noindent\textbf{Model and task generalisability:} Our evaluation focuses on CNNs for image classification, representing a standard edge deployment workload. We acknowledge that architectures with extreme outlier activations, such as Transformers, or tasks in different domains, possess different sensitivity profiles. Assessing the transferability of \toolname to these workloads and its scalability to large models remains a direction for future work.

\subsection{Construct Validity} 

\noindent\textbf{Proxy reliability:} Static proxies are susceptible to runtime anomalies and backend-specific compiler optimisations. We address this by coupling the search with an online logarithmic calibrator, dynamically aligning the surrogate models with observed hardware behaviour across successive generations to prevent the search from being misled by proxy inaccuracies. 

\noindent\textbf{Calibration set representativeness:} We use a 256-image subset for calibration and entropy profiling. This aligns with standard post-training quantisation practices and is deliberately restricted to enforce strict data isolation and prevent quantisation over-fitting.

\subsection{Conclusion Validity} 

\noindent\textbf{Measurement noise:} Hardware execution, particularly on heavily parallel targets like the \targetServerGPU, is subject to dispatch overheads and measurement noise. We mitigated this during our final validation by using extended warm-up iterations, executing multiple timed runs under strict thermal controls, and averaging the results. 

\noindent\textbf{Search stochasticity:} Due to the stochastic nature of evolutionary algorithms, all searches were repeated 5 times with varying random seeds. Results are reported as mean and variance to ensure statistical reliability.
\section{Related Work} \label{sec:relatedwork}

\begin{table*}[t] 
\centering 
\footnotesize 
\setlength{\tabcolsep}{4pt} 
\renewcommand{\arraystretch}{1.08} 
\caption{Comparison with representative mixed-precision and deployment-aware optimisation approaches. \emph{Direct feedback} denotes measurement on the target device or runtime. \emph{surrogate} includes sensitivity metrics, hardware proxies, and learned surrogates used to structure search.} 
\label{tab:related-comparison} 
\begin{tabularx}{\textwidth}{lLLccc} 
\toprule 
\textbf{Method} & \textbf{Deployment setting} & \textbf{Optimisation scheme} & \textbf{\shortstack{Direct feedback}} & \textbf{\shortstack{surrogate}} & \textbf{\shortstack{Compiler integrated}} \\ 
\midrule

    HAQ~\cite{wangHAQHardwareAwareAutomated2019}
    & Specialised accelerators
    & Reinforcement learning + HIL
    & \cmark & \xmark & \xmark \\

    SHQ~\cite{huSingleStepHardwareAwareNeural2026}
    & FPGA
    & One-shot resource model + GA
    & \xmark & \cmark & \xmark \\

    SeQTO~\cite{louloudakisSelectiveQuantizationTuner2025}
    & Commodity devices via \ONNX{}
    & Pareto search + profiling
    & \cmark & \xmark & \xmark \\

    Balaskas et al.~\cite{balaskasHardwareAwareDNNCompression2024}
    & Edge deployment
    & Pruning + MPQ + energy model
    & \xmark & \cmark & \xmark \\

    AIQ~\cite{singhArithmeticIntensityAwareQuantization2025}
    & Memory-bound inference
    & AI-aware search
    & \xmark & \cmark & \xmark \\

    HAWQ-V3~\cite{yaoHAWQV3DyadicNeural2021}
    & Integer-only deployment
    & Hessian-guided ILP
    & \xmark & \cmark & \xmark \\

    InfoQ~\cite{akbulutInfoQMixedPrecisionQuantization2026}
    & Resource-bounded MPQ
    & Global information flow + ILP
    & \xmark & \cmark & \xmark \\

    LCPAQ~\cite{chenAdaptiveQuantizationMixedPrecision2024}
    & Hardware-constrained MPQ
    & Hessian + ILP + proxy NAS
    & \xmark & \cmark & \xmark \\

    GMPQ-TE~\cite{liEfficientGeneralizableMixedPrecision2025}
    & Generalisable MPQ
    & Topological entropy + LP
    & \xmark & \cmark & \xmark \\

    EvoQ~\cite{yuanEvoQMixedPrecision2020}
    & General MPQ
    & Evolutionary search + sensitivity
    & \xmark & \cmark & \xmark \\

    HMQAT~\cite{huangHessianbasedMixedprecisionQuantization2025}
    & General MPQ
    & Hessian-based Pareto optimisation
    & \xmark & \cmark & \xmark \\

    AutoQRA~\cite{zhouAutoQRAJointOptimization2026}
    & LLM adaptation
    & Surrogate-guided search
    & \xmark & \cmark & \xmark \\

    \textbf{\toolname{}}
    & Commodity devices in \MLIR{}
    & MOGA + hardware proxy + HIL
    & \cmark & \cmark & \cmark \\
    \bottomrule
\end{tabularx}
\end{table*}

Table~\ref{tab:related-comparison} separates deployment settings, optimisation schemes, and the sources of guidance used during the search. This makes it easier to distinguish commodity-device systems such as SeQTO from accelerator- or FPGA-orientated approaches such as HAQ and SHQ, and from model-centric mixed-precision methods such as HAWQ-V3, InfoQ, LCPAQ, GMPQ-TE, EvoQ, and HMQAT. \toolname{} combines commodity-device deployment with compiler-integrated search and pre-HIL screening.

\textbf{Compiler infrastructures for machine learning deployment:} Modern ML deployment pipelines remain fragmented across model conversion, quantisation, compiler lowering, and backend runtime support. \MLIR{} addresses part of this fragmentation by providing a reusable multi-level compiler infrastructure for heterogeneous workloads and hardware targets \cite{lattnerMLIRCompilerInfrastructure2020}. Frameworks such as TinyIREE show how \MLIR{}-based lowering through dialects such as \TOSA{} and \Linalg{} can target desktop, mobile, and embedded devices within a unified execution environment \cite{liuTinyIREEMLExecution2022}. However, these pipelines generally treat quantisation as an external preprocessing step and expect quantised models to be imported from front-end tools. They therefore provide limited support for generating or searching mixed-precision candidates inside the compiler itself. \toolname{} targets this gap by generating, lowering, and evaluating mixed-precision programs within a single pipeline.

\textbf{Deployment-aware quantisation and model compression:} Several methods adapt quantisation to deployment constraints, but differ substantially in their target setting and source of hardware guidance. HAQ uses reinforcement learning driven by direct latency and energy feedback, but is aimed at specialised accelerator settings rather than common deployment stacks \cite{wangHAQHardwareAwareAutomated2019}. SHQ estimates hardware resource usage in a single step before deployment, but focuses on FPGA-based full-pipeline accelerators \cite{huSingleStepHardwareAwareNeural2026}. Balaskas et al.~combine pruning and mixed-precision quantisation under a hardware-aware energy model for edge deployment \cite{balaskasHardwareAwareDNNCompression2024}. AIQ optimises per-layer bit-widths to maximise arithmetic intensity, targeting memory-bound inference through an analytical performance objective rather than direct device profiling \cite{singhArithmeticIntensityAwareQuantization2025}. Closest to our deployment setting, SeQTO performs selective quantisation and on-device profiling of \ONNX{} models across CPU, GPU, and mobile targets to identify Pareto-optimal candidates \cite{louloudakisSelectiveQuantizationTuner2025}. \toolname{} shares this commodity-device focus, but differs in where and how search is performed: it moves candidate generation into the compiler IR and applies low-cost hardware filtering before expensive HIL execution.

\textbf{Constraint-based and sensitivity-guided mixed-precision quantisation:} A second strand of work reduces MPQ search cost by replacing repeated full evaluations with analytical constraints or layer-sensitivity signals. HAWQ-V3 uses Hessian information and integer linear programming to allocate bit-widths under integer-only deployment constraints \cite{yaoHAWQV3DyadicNeural2021}. EvoQ and HMQAT likewise use sensitivity-guided search, relying on evolutionary exploration or Pareto optimisation to navigate the MPQ space \cite{yuanEvoQMixedPrecision2020, huangHessianbasedMixedprecisionQuantization2025}. InfoQ replaces local sensitivity heuristics with a global information-flow metric based on mutual information, again solving the final allocation through ILP under resource budgets \cite{akbulutInfoQMixedPrecisionQuantization2026}. LCPAQ combines Hessian-based sensitivity, ILP-based hardware constraints, and a low-cost proxy NAS module to reduce search effort \cite{chenAdaptiveQuantizationMixedPrecision2024} -- specifically -- relying on a learned Multi-Layer Perceptron proxy to simulate relative accuracy offline. GMPQ-TE uses topological entropy to derive a single-pass linear programme that yields generalisable quantisation policies across datasets \cite{liEfficientGeneralizableMixedPrecision2025}. These methods show the value of constraints and low-cost guidance, but they remain largely model-centric: they do not account for compiler lowering decisions, neighbouring-operator transition costs, backend operator availability, or runtime execution through a deployment stack such as \IREE{}.

\textbf{Proxy-guided optimisation and search efficiency:} Mixed-precision search is combinatorial, so using cheap guidance signals to reduce expensive evaluations is a natural strategy. AutoQRA uses surrogate models to discard poor configurations before expensive downstream optimisation in the joint search over mixed precision and low-rank adaptation settings \cite{zhouAutoQRAJointOptimization2026}. Fundamentally, using surrogate models to assist Multi-Objective Evolutionary Algorithms (MOEAs) is a well-established technique for mitigating such expensive optimisation problems~\cite{lim2009generalizing, chugh2019surrogate, li2022classification}. More broadly, Search-Based Software Engineering has long used approximate fitness signals to avoid unnecessary full evaluations, for example, in allocator optimisation \cite{Dakhama2025GreenMalloc}, while training-free neural architecture search combines multiple low-cost signals to steer exploration efficiently \cite{leeAZNASAssemblingZeroCost2024}. While other frameworks utilise surrogate-guided search to optimise tensor programs \cite{chen2018tvm, chen2018learning} or apply reinforcement learning to MLIR's Linalg dialect \cite{tirichine2024reinforcement}, they generally do not natively support mixed-precision generation within the IR.\toolname{} follows the same principle as these approaches, providing compiler-integrated MPQ: a lightweight hardware cost model filters weak candidates before full compilation and \IREE{} execution on the target device.
\section{Conclusion and Future Work}
\label{sec:conclusion}

We presented \toolname{}, a hardware-aware mixed-precision quantisation infrastructure that shifts the quantisation process directly into the \MLIR{} \TOSA{} compiler pipeline. For software engineers and ML practitioners, this integration provides a unified deployment layer that resolves the severe fragmentation between frontend frameworks and backend compilers~\cite{jajal2024interoperability,wang2023compatibility,daoudi2025neural}. By generating and optimising configurations entirely within the intermediate representation, developers avoid brittle toolchain conversions and are guaranteed that their quantised profiles are structurally compatible with their deployment system. 

While our dual-surrogate, online-calibrated search introduces a higher upfront computational cost compared to some hardware-agnostic analytical methods, this investment is essential for resource-constrained environments. When runtime latency is a strict deployment constraint, the overhead of hardware-in-the-loop (HIL) feedback is quickly recovered across a large deployment.

Future work will expand \toolname{} across three primary vectors. First, we plan to refine the accuracy and latency surrogates to further minimise the HIL evaluation budget. Second, we aim to extend our sensitivity profiling and search mechanism to support highly dynamic architectures such as Transformers. Finally, we will explore deep co-optimisation at the compiler level, where downstream compilation passes (\eg, operator tiling, and vectorisation) are dynamically tuned in tandem with the automatically discovered quantisation configurations, providing even greater end-to-end deployment efficiency, and developer ease.

\bibliographystyle{IEEEtran}
\bibliography{references}

\end{document}